\documentclass[aps,prd,twocolumn,showpacs,superscriptaddress]{revtex4}
\usepackage{graphics}

\newcommand{\ba}{\begin{array}}
\newcommand{\ea}{\end{array}}
\newcommand{\beqa}{\begin{eqnarray}}
\newcommand{\eeqa}{\end{eqnarray}}

\begin{document}

\title{Hierarchical radiative masses in a supersymmetric three-family model 
without Higgsinos}
\author{Luis A. S\'anchez}
\affiliation{Escuela de F\'\i sica, Universidad Nacional de Colombia,
A.A. 3840, Medell\'\i n, Colombia}
\author{William A. Ponce}
\affiliation{Instituto de  F\'\i sica, Universidad de Antioquia,
A.A. 1226, Medell\'\i n, Colombia}
\author{Jes\'us M. Mira} 
\affiliation{Instituto de F\'\i sica, Universidad de Antioquia,
A.A. 1226, Medell\'\i n, Colombia}

\begin{abstract}
We study the scalar potential and the mass spectrum of the supersymmetric
extension of a three-family model based on the local gauge group
$SU(3)_C\otimes SU(3)_L\otimes U(1)_X$, with anomalies canceled among the
three families in a nontrivial fashion. In this model the
slepton multiplets play the role of the Higgs scalars and no Higgsinos are
required, with the consequence that the sneutrino, the selectron and six
other sleptons play the role of the Goldstone bosons of the theory. By 
introducing an Abelian anomaly-free discrete symmetry and aligning the 
vacuum in a convenient way, we get a consistent mass spectrum for the 
scalars and for the spin 1/2 quarks and charged leptons, where only the 
top and charm quarks and the tau lepton acquire tree level masses while the remaining ordinary charged fermions acquire radiative hierarchical masses.  
\end{abstract}

\maketitle

\section{\label{sec:intr}Introduction} 
Among the unsolved questions of the standard model (SM), which is a theory
based on the local gauge group $SU(3)_c\otimes SU(2)_L\otimes U(1)_Y$
\cite{sm}, the elucidation of the nature of the electroweak symmetry
breaking remains one of the most challenging issues. If the electroweak
symmetry is spontaneously broken by Higgs scalars, the determination of
the value of the Higgs mass $M_H$ in the context of the SM becomes a key
ingredient. By direct search LEP-II has set an experimental lower bound
for a neutral scalar, member of a pure $SU(2)_L$ doublet, of 114.4
GeV \cite{lep2} .

Today, supersymmetry (SUSY) is considered as the leading candidate for new
physics. Even though SUSY does not solve all open questions, it has
attractive features, the most important one being that it protects the
electroweak scale from destabilizing divergences, that is, SUSY provides
an answer as to why the scalars remain massless down to the electroweak scale
where there is no symmetry protecting them (the ``hierarchy problem").  
This has motivated the construction of the minimal supersymmetric standard
model (MSSM) \cite{mssm}, the supersymmetric extension of the SM, that is
defined by the minimal field content and minimal superpotential necessary
to account for the known Yukawa mass terms of the SM. At present, however,
there is no experimental evidence that Nature is supersymmetric, and the
only experimental fact that points toward a beyond the SM structure lies
in the neutrino sector, and even there the results are not final yet. So, a
reasonable approach is to depart from the SM as little as possible,
allowing for some room for neutrino oscillations \cite{neutrinos}.

In that direction, over the last decade several variants of the SM extension based on the local gauge group $SU(3)_c\otimes SU(3)_L\otimes U(1)_X$ (hereafter called the 3-3-1 structure), where anomalies cancel by an interplay between the three
families, have received special attention. In some of them the three known
left-handed lepton components for each family are associated to three
$SU(3)_L$ triplets as $(\nu_l, l^-, l^+)^T_L$, where $l^+_L$ is related to
the right-handed isospin singlet of the charged lepton $l_L^-$ in the SM
\cite{fp} . In other models the three $SU(3)_L$ lepton triplets are of the
form $(\nu_l,l^-,\nu_l^c)^T_L$ where $\nu_{lL}^c$ is related to the
right-handed component of the neutrino field $\nu_{lL}$ \cite{long,
ponce}. There are also models in the literature with $SU(3)_L$ lepton
triplets of the form $(l^-,\nu_l,L^-)_L^T$, where $L_L^-$ is an exotic
charged lepton with electric charge $-1$ \cite{ponce, ozer}. In the first
model anomaly cancellation requires quarks with exotic electric charges
$-4/3$ and 5/3 which in turn imply double charged gauge and Higgs bosons,
while in the other models the exotic particles have only ordinary electric
charges.

As it is shown in Ref.~\cite{ponce}, there are just a few different 
non-supersymmetric three-family models based on the 3-3-1 local gauge
structure which are free of chiral anomalies and do not include particles
with exotic electric charges. These models share in common not only the
same gauge boson sector, but also the same scalar sector. In this paper we
are going to construct a consistent SUSY extension of a three-family model
which is the simplest one with regard to the three-family 3-3-1 models
introduced in Ref.~\cite{ponce}.

Our main motivation lies in the 3-3-1 SUSY one-family model presented in
Ref.~\cite{331susy}, in which the left-handed lepton triplets and the
Higgs scalars needed to break the symmetry down to $SU(3)_c \otimes
U(1)_Q$, have the same quantum numbers under the gauge group, and may
play the role of the superpartners of each other. As a result, in the one-family model several consequences follow \cite{331susy}: first, the reduction of the number of free parameters as compared to supersymmetric versions of other 3-3-1 models in the literature \cite{ma}; second, the result that the sneutrino, selectron and six other sleptons do not acquire masses in the context of the model constructed playing the role of the Goldstone bosons; third, the absence of the $\mu$-problem, in the sense that the $\mu$-term is absent at the tree level, arising only as a result of the symmetry breaking, and fourth, the existence of light CP-odd scalars which may have escaped experimental detection.

Our aim in this paper is to explore the possibility to construct a
realistic three-family 3-3-1 SUSY model as far as the particle mass
spectrum is concerned; the price we have to pay is an alignment of the
vacuum state and the introduction of a discrete symmetry, as we will show
in due course. The paper is organized as follows:  in Sec.~\ref{sec:sec2} we
briefly review the non-supersymmetric version of the model, in
Sec.~\ref{sec:sec3} we comment on its supersymmetric extension and
calculate the superpotential, in Sec.~\ref{sec:sec4} we calculate the mass
spectrum (excluding the squark sector), and in Sec.~\ref{sec:sec5} we
present our conclusions.

\section{\label{sec:sec2}The non-supersymmetric model} 
The model we are going to supersymmetrize is based on
the local gauge group $SU(3)_c\otimes SU(3)_L\otimes U(1)_X$. It has 17
gauge bosons: one gauge field $B^\mu$ associated with $U(1)_X$, the 8
gluon fields $G^\mu$ associated with $SU(3)_c$ which remain massless after
breaking the symmetry, and another 8 gauge fields associated with
$SU(3)_L$ and that we write for convenience as \cite{ponce}

\[{1\over 2}\lambda_\alpha A^\mu_\alpha={1\over \sqrt{2}}\left(
\begin{array}{ccc}D^\mu_1 & W^{+\mu} & K^{+\mu} \\ W^{-\mu} & D^\mu_2 &
K^{0\mu} \\ K^{-\mu} & \bar{K}^{0\mu} & D^\mu_3 \end{array}\right), \]
where $D^\mu_1=A_3^\mu/\sqrt{2}+A_8^\mu/\sqrt{6},\;
D^\mu_2=-A_3^\mu/\sqrt{2}+A_8^\mu/\sqrt{6}$, and
$D^\mu_3=-2A_8^\mu/\sqrt{6}$. $\lambda_i, \; i=1,2,...,8$, are the eight
Gell-Mann matrices normalized as $Tr(\lambda_i\lambda_j)  
=2\delta_{ij}$.

The charge operator associated with the unbroken gauge symmetry $U(1)_Q$ 
is given by
\begin{equation}\label{chop}
Q=T_{3L}+\frac{T_{8L}}{\sqrt{3}}+XI_3,
\end{equation}
where $T_{iL}=\lambda_{iL}/2$, $I_3=Dg.(1,1,1)$ is the diagonal $3\times 3$ unit matrix, and the $X$ values are related to the $U(1)_X$ hypercharge and are fixed by anomaly cancellation. Eq.~(\ref{chop}) is a particular case 
of the most general expresion for the electric charge generator in $SU(3)_c\otimes SU(3)_L\otimes U(1)_X$ given by: $Q=a T_{3L}/2+ 2b T_{8L}/\sqrt{3}+XI_3$, where $a$ and $b$ are free parameters, and 
corresponds to the choice $a=1$ (in order for the weak isospin to be contained in $SU(3)_L$) and $b=1/2$ (so that the model does not contain exotic electric charges) \cite{ponce}. 

The sine of the electroweak mixing angle is given by 
$S_W^2=3g_1^2/(3g_3^2+4g_1^2)$ where $g_1$ and $g_3$ are the coupling 
constants of $U(1)_X$ and $SU(3)_L$ respectively, and the photon field is 
given by
\begin{equation}\label{foton}
A_0^\mu=S_WA_3^\mu+C_W\left[\frac{T_W}{\sqrt{3}}A_8^\mu + 
\sqrt{(1-T^2_W/3)}B^\mu\right],
\end{equation}
where $C_W$ and $T_W$ are the cosine and tangent of the electroweak mixing 
angle, respectively. 

There are two neutral currents in the model which are defined as
\begin{eqnarray}\nonumber
Z_0^\mu&=&C_WA_3^\mu-S_W\left[\frac{T_W}{\sqrt{3}}A_8^\mu + 
\sqrt{(1-T^2_W/3)}B^\mu\right], \\ \label{zetas}
Z_0^{\prime\mu}&=&-\sqrt{(1-T^2_W/3)}A_8^\mu+\frac{T_W}{\sqrt{3}}B^\mu,
\end{eqnarray}
where $Z^\mu_0$ coincides with the weak neutral current of the SM. Using  
Eqs.~(\ref{foton}) and (\ref{zetas}) we may read the gauge boson $Y^\mu$ 
associated with the $U(1)_Y$ hypercharge in the SM
\[Y^\mu=\frac{T_W}{\sqrt{3}}A_8^\mu + 
\sqrt{(1-T^2_W/3)}B^\mu. \]

The quark content for the three families is the following \cite{ponce}:  
$Q_{iL}=(u_i,d_i,D_i)_L^T\sim(3,3,0),\;i=2,3,$ for two families, where
$D_{iL}$ are two exotic quarks of electric charge $-1/3$ (the numbers
inside the parentheses stand for the $[SU(3)_c,SU(3)_L,U(1)_X]$ quantum
numbers in that order); $Q_{1L}=(d_1,u_1,U)^T_L\sim (3,3^*,1/3)$, where
$U_L$ is an exotic quark of electric charge $2/3$. The right-handed quarks
are: $u^{c}_{aL}\sim (3^*,1,-2/3)$, $d^{c}_{aL}\sim (3^*,1,1/3)$, with
$a=1,2,3,$ a family index, $D^{c}_{iL}\sim (3^*,1,1/3)$, and $U^c_L\sim
(3^*,1,-2/3)$.

The lepton content is given by the three anti-triplets $L_{\alpha L} =
(\alpha^-,\nu_\alpha^0,N_\alpha^0)^T_L\sim (1,3^*,-1/3)$, the three
singlets $\alpha^+_{L}\sim(1,1,1)$, $\alpha = e, \mu, \tau$, and the
vectorlike structure (vectorlike with respect to the 3-3-1 gauge group)  
$L_{4L}=(N_4^0,E_4^+,E_5^{+})^T_L\sim (1,3^*,2/3)$, and
$L_{5L}=(N_5^0,E_4^-,E_5^{-})^T_L\sim (1,3,-2/3)$; where $N^0_s,\;  
s=e, \mu, \tau, 4,5$, are five neutral Weyl states, and $E^-_\eta, \eta =
4,5 $ are two exotic electrons.
 
With the former quantum numbers it is just a matter of counting to check
that the model is free of the following chiral anomalies:  $[SU(3)_c]^3\;
(SU(3)_c$ is vectorlike); $[SU(3)_L]^3$ (seven triplets and seven
anti-triplets), $[SU(3)_c]^2U(1)_X; \; [SU(3)_L]^2U(1)_X ;
\;[grav]^2U(1)_X$ (the so called gravitational anomaly \cite{del}) and
$[U(1)_X]^3$.

For this model the minimal scalar sector, able both to break the symmetry
and to give, at the same time, masses to the
fermion fields, is given by \cite{pgs}:  $\chi_1^T=(\chi^-_1, \chi^0_1,
\chi^{'0}_1) \sim (1,3^*,-1/3)$, and $\chi_2^T=(\chi^0_2, \chi^+_2,
\chi^{'+}_2) \sim (1,3^*,2/3)$, with vacuum expectation values (VEV)
given by $\langle\chi_1\rangle^T=(0,v_1,V)$ and
$\langle\chi_2\rangle^T=(v_2,0,0)$, with the hierarchy 
$V \gg v_1 ,\; v_2$. These VEV break the symmetry 
\[SU(3)_c\otimes SU(3)_L\otimes U(1)_Y\longrightarrow SU(3)_c\otimes 
U(1)_Q\] 
in one single step and so, the SM can not be considered as an effective 
theory of this particular 3-3-1 gauge structure. 

This model, even though scketched in Refs.~\cite{ponce, pgs} where it was 
named Model E, has not been studied in the literature as far as 
we know. (A related model without the vector-like structure 
$L_{4L}\oplus L_{5L}$ and with a scalar sector of three triplets 
instead of two, has been partially analyzed in Ref.~\cite{long}.)

Notice that in the nonsupersymmetric model, universality for the known
leptons in the three families is present at the tree level in the weak
basis, up to mixing with the exotic fields. Since the mass scale of the new neutral gauge boson $Z'$ and of the exotic particles is of the order of $V$, this mixing will suppress tree-level flavor changing neutral currents (FCNC) effects in the lepton sector. For the quarks, instead, one family transform differently from the other two and, as a result, there can be potentially large FCNC in the hadronic sector. Since it is not our goal to discuss this issue here, the reader is referred to the recent study presented in Ref.~\cite{sher}. Let us, notwithstanding, point out that the present model is associated to the one called Model B in Ref.~\cite{sher}, for which the constraints imposed by flavor changing phenomenology in the quark sector are not so severe as for other 3-3-1 models.

\section{\label{sec:sec3}The supersymmetric extension} 
When supersymmetry is introduced in the SM, the entire spectrum of
particles is doubled as we must introduce the superpartners of the known
fields. Also, two scalar doublets $\phi_u$ and $\phi_d$ must be used in
order to cancel the triangle anomalies; then the superfields
${\hat\phi}_u$ and ${\hat\phi}_d$, related to the two scalars, may couple
via a term of the form $\mu {\hat\phi}_u{\hat\phi}_d$ which is gauge and
supersymmetric invariant, and thus the natural value for $\mu$ is expected
to be much larger than the electroweak and supersymmetry breaking scales.
This is the so-called $\mu$ problem.

However, in a non supersymmetric model as the one presented in the former
section, in which the Higgs fields transform as some of the lepton fields under the symmetry group, the SUSY extension can be constructed with the scalar and the lepton fields acting as superpartners of each other, ending up with a SUSY model without Higgsinos \cite{331susy}, which is automatically free of chiral anomalies.

For three families we thus have the following chiral superfields:
$\hat{Q}_a$, $\hat{u}_a$, $\hat{d}_a$, $\hat{D}_i$, $\hat{U}$,
$\hat{L}_a$, $\hat{e}_a$, and $\hat{L}_\eta$, plus 
gauge bosons and gauginos, where $a=1,2,3$ is a family index, $i=1,2,$ and $\eta =4,5$. The identification of the gauge bosons eigenstates in the SUSY extension follows the non-SUSY version, as it will be shown in Sec.~\ref{sec:sec41}.

\subsection{\label{sec:sec31}The Superpotential}
Let us now write the most general $SU(3)_c\otimes SU(3)_L\otimes U(1)_X$ 
invariant superpotential for the model
\beqa \label{superpotential} 
W&=&h^u_{ia}\hat{Q}_i\hat{u}_a\hat{L}_4 +h^U_i\hat{Q}_i\hat{U}\hat{L}_4+
h^d_{iab}\hat{Q}_i\hat{d}_a\hat{L}_b \nonumber\\
& & +h^D_{ija}\hat{Q}_i\hat{D}_j\hat{L}_a 
+h^{d\prime}_a\hat{Q}_1\hat{d}_a\hat{L}_5 
+ h^{\prime D}_i\hat{Q}_1\hat{D}_i\hat{L}_5 \nonumber\\
& & + h^{e}_{ab}\hat{L}_a\hat{e}_b\hat{L}_5
+ \frac{1}{2}\lambda_{ab}\hat{L}_a\hat{L}_b\hat{L}_4 
+ \mu \hat{L}_4\hat{L}_5 \nonumber\\
& & + \lambda^{(1)}_{abi}\hat{u}_a\hat{d}_b\hat{D}_i
+ \lambda^{(2)}_{ai}\hat{U}\hat{d}_a\hat{D}_i
+ \lambda^{(3)}_{ijk}\hat{Q}_i\hat{Q}_j\hat{Q}_k \nonumber\\ 
& & + \lambda^{(4)}_{abc}\hat{u}_a\hat{d}_b\hat{d}_c +
\lambda^{(5)}_{aij}\hat{u}_a\hat{D}_i\hat{D}_j +
\lambda^{(6)}_{ab}\hat{U}\hat{d}_a\hat{d}_b \nonumber\\
& & + \lambda^{(7)}_{ij}\hat{U}\hat{D}_i\hat{D}_j\,,
\eeqa 
where summation over repeated indexes is understood, and the chirality,
color and isospin indexes have been omitted.

The $\hat{u}\hat{d}\hat{D}$, $\hat{U}\hat{d}\hat{D}$,
$\hat{U}\hat{d}\hat{d}$, $\hat{U}\hat{D}\hat{D}$, $\hat{u}\hat{d}\hat{d}$,
$\hat{u}\hat{D}\hat{D}$, and $\hat{Q}\hat{Q}\hat{Q}$ terms violate
baryon number and can lead to rapid proton decay. We may forbid
these interactions by introducing an anomaly free discrete $Z_2$ 
symmetry \cite{KW} with the following assignments of $Z_2$ charge $q$ 
\begin{eqnarray}\nonumber \label{z2}
q(\hat{Q}_a, \hat{u}_a,\hat{U},\hat{D}_i,\hat{d}_a,\hat{\mu},\hat{L}_\mu)= 
1, \nonumber \\ 
q(\hat{L}_e,\hat{L}_\tau,\hat{e}, \hat{\tau}, \hat{L}_4, 
\hat{L}_5)=0,
\end{eqnarray}
where we have used $\hat{e}_1 \equiv \hat{e}$, $\hat{e}_2 \equiv \hat{\mu}$, 
$\hat{e}_3 \equiv \hat{\tau}$, $\hat{L}_1 \equiv \hat{L}_e$, 
$\hat{L}_2 \equiv \hat{L}_\mu$, and $\hat{L}_3 \equiv \hat{L}_\tau$. 
This is just one among several anomaly-free discrete symmetries 
available. This symmetry protects the model from a too fast proton decay, 
but the superpotential still contains operators inducing lepton number 
violation, which is desirable if we want to describe Majorana masses for the neutrinos in the model.

The $Z_2$ symmetry also forbids some undesirable mass terms for the spin
$1/2$ fermions which complicate unnecessarily several mass matrices. But
notice the presence of a $\mu$ term in the superpotential that, as we will
show in a moment is convenient to keep, in order to have a consistent mass
spectrum. So, contrary to the model in Ref.~\cite{331susy}, this model has
a $\mu$-term coming from the existence of the vectorlike structure
$\hat{L}_{4L}\oplus\hat{L}_{5L}$.

\subsection{\label{sec:sec32}The scalar potential} 
The scalar potential is written as 
\begin{equation} \label{potesc}
V_{SP}=V_F+V_D+V_{\hbox{soft}},
\end{equation} 
where the first two terms come from the exact SUSY sector, while the last
one is the sector of the theory that breaks SUSY explicitly.

We now display $V_F$ in Eq.~(\ref{potesc}), before implementing the 
discrete $Z_2$ symmetry

\beqa \nonumber \label{Vf} 
V_F&=&
\sum^3_{a=1}\left |\frac{\partial W}{\partial\tilde{L}_a}
\right |^2 +
\sum_{\eta=4}^5\left |\frac{\partial W}{\partial\tilde{L}_\eta}
\right |^2 \\ \nonumber
&=&(\lambda^\dagger \lambda)_{ab}\left \{
(\tilde{L}^{\dagger}_a\tilde{L}_b)|\tilde{L}_4|^2 -
(\tilde{L}^{\dagger}_a\tilde{L}_4)
(\tilde{L}^{\dagger}_4\tilde{L}_b)\right \} \\ \nonumber
&+&\tilde{e}^{\dagger}_a H_{ab}^e 
\tilde{e}_b|\tilde{L}_5|^2 +(\tilde{L}^{T}_a \tilde{L}_5)
(h^eh^{e\dagger})_{ab}\tilde{L}^{T}_b \tilde{L}_5\\ \nonumber 
&+&\left\{(\lambda^\dagger 
h^e)_{ab}\tilde{L}^{*}_a \cdot
(\tilde{L}^{*}_4\times \tilde{L}_5)\tilde{e}_b+
c.c.\right\} \\ \nonumber
&+&|\mu|^2\left( |\tilde{L}_4|^2 +|\tilde{L}_5|^2\right) 
+\left\{\mu^*h_{a b}^e(\tilde{L}^{\dagger}_4 \tilde{L}_a)
\tilde{e}_b+ c.c.\right\} \\ \nonumber
&+&\frac{1}{4}\lambda_{ab}{\lambda_{cd}^*}
\left \{(\tilde{L}^{\dagger}_c\tilde{L}_a)
(\tilde{L}^{\dagger}_d\tilde{L}_b)
-(\tilde{L}^{\dagger}_d\tilde{L}_a)
(\tilde{L}^{\dagger}_c\tilde{L}_b)\right\} \\ \nonumber
&+ &h_{a b}^eh_{c d}^{e*}(\tilde{L}^{\dagger}_c
\tilde{L}_a)\tilde{e}_b\tilde{e}^{*}_d \\ 
&+&\left\{\frac{1}{2}\mu^*\lambda_{a b}\tilde{L}_a \cdot
(\tilde{L}_b \times \tilde{L}^{*}_5)+ c.c.\right\}, 
\eeqa
where $H^e=(h^{e\dagger}h^e)$ is an hermitian $3\times 3$ matrix, and 
$\tilde{L}_a\cdot (\tilde{L}_b\times \tilde{L}_c)$ is a 
triple scalar product in the tridimensional lineal representation of 
$SU(3)_L$.

When the $Z_2$ symmetry is introduced $V_F$ gets reduced to the 
expression
\beqa \nonumber \label{vfz}
V_F&=&\left|\lambda \tilde{L}_3\times\tilde{L}_4+h_{11}^e
\tilde{e}\tilde{L}_5+h_{13}^e\tilde{\tau}\tilde{L}_5\right|^2 \nonumber \\
&+&
\left |-\lambda \tilde{L}_1\times\tilde{L}_4+h_{33}^e
\tilde{\tau}\tilde{L}_5+h_{31}^e\tilde{e}\tilde{L}_5\right|^2\nonumber \\
&+&\left|\lambda \tilde{L}_1\times\tilde{L}_3+\mu\tilde{L}_5\right|^2 
\nonumber \\
&+&|h^e_{22}|^2|\tilde{L}_{5}|^2|\tilde{\mu}|^2+
|h^e_{22}|^2|\tilde{L}^{T}_2\tilde{L}_5|^2\nonumber \\
&+&\left|h^e_{11}\tilde{L}^{T}_1\tilde{L}_5+
h^e_{31}\tilde{L}^{T}_3\tilde{L}_5\right|^2\nonumber \\
&+&\left|h^e_{13}\tilde{L}^{T}_1\tilde{L}_5+
h^e_{33}\tilde{L}^{T}_3\tilde{L}_5\right|^2
\nonumber \\
&+&|\mu \tilde{L}_4+ h^e_{22} \tilde{\mu}\tilde{L}_2+(h_{11}^e
\tilde{e}+h^e_{13}\tilde{\tau})\tilde{L}_1\nonumber \\ 
& &+(h_{31}^e\tilde{e}+h^e_{33}\tilde{\tau}) 
\tilde{L}_3|^2, 
\eeqa
where $\lambda= \lambda_{13}=-\lambda_{31}\equiv \lambda_{e\tau}$ is the 
only $\lambda_{ab}$ parameter which survives. This form of $V_F$ is crucial 
for the analysis that follows.

For the second term in $V_{SP}$ we have
\begin{eqnarray}\nonumber
V_D&=&\frac{1}{2}D^\alpha D^\alpha+\frac{1}{2}D^2 \\ \nonumber
& = & \frac{1}{4}g_3^2
\Bigg\{\sum_{\alpha,\beta }^{4}\left(|\tilde{L}^{\dagger}_{\alpha}
\tilde{L}_{\beta}|^2-\frac{1}{3}|\tilde{L}_{\alpha}|^2
|\tilde{L}_{\beta}|^2\right)\nonumber \qquad \qquad \\ \nonumber
& &+\sum_{\alpha }^{4}\left(2|\tilde{L}^{\dagger}_{\alpha}
\tilde{L}_{5}|^2-\frac{2}{3}|\tilde{L}_{\alpha}|^2
|\tilde{L}_{5}|^2\right)+\frac{2}{3}|\tilde{L}_{5}|^4
\Bigg\} \\ \nonumber
& &+\frac{1}{18}g_1^2\Bigg\{
\left(\sum_{a}^{3}|\tilde{L}_a|^2\right)^2
+4|\tilde{L}_{4}|^4+4|\tilde{L}_{5}|^4 \\ \nonumber
& &-8|\tilde{L}_{4}|^2|\tilde{L}_{5}|^2
+4\sum_{a}^{3}|\tilde{L}_a|^2|\tilde{L}_{5}|^2 \\ \label{Vd}
& &-4\sum_{a}^{3}|\tilde{L}_a|^2|\tilde{L}_{4}|^2
\Bigg\}\,. 
\end{eqnarray}

The soft SUSY-breaking scalar potential is
\beqa \nonumber \label{Vsoft}
V_{\hbox{soft}}&=&
m_{ab}^2\hbox{Re}(\tilde{L}^{\dagger}_a\tilde{L}_b) 
+m_{4}^2|\tilde{L}_4|^2 + m_{5}^2|\tilde{L}_5|^2 \\ \nonumber
&+& m_{45}^2\hbox{Re}(\tilde{L}^{T}_4\tilde{L}_5) 
+\hbox{Re}(h^{e\prime}_{ab}\tilde{L}^{T}_a  
\tilde{L}_5\tilde{e}_b) \\ \nonumber
&+& \frac{\epsilon^{ijk}}{2}\hbox{Re}
\,(\lambda_{ab}^\prime\tilde{L}_{ai}
\tilde{L}_{bj}\tilde{L}_{4k}) \nonumber \\  
&+& \frac{M_1}{2}\tilde{B}^0\tilde{B}^0 
+ \frac{M_2}{2}\sum_{a=1}^8 \tilde{A}_a\tilde{A}^a + \dots ,
\eeqa
where $M_1$ is the soft mass parameter of the $U(1)_X$ gaugino and $M_2$
refers to the soft mass parameter of the $SU(3)_L$ gauginos.

\subsection{\label{sec:sec33}The vacuum}
In this model we do not introduce Higgs scalars as it is done for example 
in the MSSM. The duty of the spontaneous symmetry breaking is assigned 
to the neutral sleptons which are present in the chiral supermultiplets 
$\hat{L}_\alpha$ and $\hat{L}_\eta$ for $\alpha = e, \mu, \tau = 1,2,3$ and 
$\eta =4,5$.

To use the most general VEV structure available, even when properly rotated, is 
a hopeless task. What we propose here is to align the vacuum in the 
following way:
$\langle\phi_e\rangle = \langle (\phi_e^-,\phi_e^0,\phi_e^{\prime 
0})\rangle = (0,0,0); \; 
\langle\phi_\mu\rangle = \langle (\phi_\mu^-,\phi_\mu^0,
\phi_\mu^{\prime 0})\rangle = (0,0,0); \;  
\langle\phi_\tau\rangle = \langle 
(\phi_\tau^-,\phi_\tau^0,\phi_\tau^{\prime 0})\rangle = (0,0,V); \;  
\langle\phi_4\rangle = \langle (\phi_4^0,\phi_4^+,
\phi_4^{\prime +})\rangle = (v,0,0)$ and 
$\langle\phi_5\rangle = \langle (\phi_5^0,\phi_5^-,
\phi_5^{\prime -})\rangle = (0,0,0)$, where we have introduced the 
notation $\phi_\alpha\equiv\tilde{L}_\alpha$ for the superpartners of 
$L_\alpha$, and $\phi_\eta 
\equiv \tilde{L}_\eta$ for the superpartners of $L_\eta$. We also will use $\phi^{\prime +}_e$, $\phi^{\prime +}_\mu$, $\phi^{\prime +}_\tau$ for the scalar superpartners of the singlets $e^+_L$, $\mu^+_L$, $\tau^+_L$, respectively.
In what follows we are going to show that for $V \gg v\sim 174$ GeV (the 
electroweak breaking scale), this alignment is enough to reproduce a 
consistent mass spectrum.

The former VEV structure allows us to break the symmetry in the way
\beqa \label{ssb} \nonumber
3-3-1&\stackrel{V}{\longrightarrow}& SU(3)_c\otimes 
SU(2)_L\otimes U(1)_Y \\
&\stackrel{v}{\longrightarrow}& SU(3)_c\otimes U(1)_Q, 
\eeqa
which in turn allows for the matching conditions $g_2=g_3$ and 
\begin{equation}
\frac{1}{g^{\prime 2}}=\frac{1}{g_1^2}+ \frac{1}{3g_3^2},
\end{equation}
where $g_2$ and $g^\prime$ are the gauge coupling constants for the 
gauge groups $SU(2)_L$ and $U(1)_Y$ of the SM, respectively.

\begin{figure}
\includegraphics{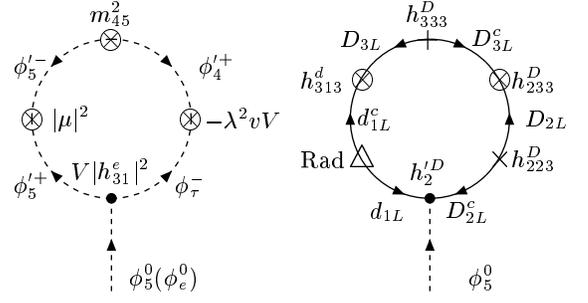}
\caption{\label{fig1}Radiatively induced VEV for 
$\phi_5^0$ and $\phi_e^0$.}
\end{figure}

A further study of the superpotential in Eq.~(\ref{superpotential}) and the scalar potential $V_F$ in Eq.~(\ref{vfz}) shows
that even if $\langle\phi_5^0\rangle=\langle\phi_e^0\rangle =0$ at the tree
level, they both develop a radiatively induced VEV different from zero, as
it is shown in Fig.~\ref{fig1}. In particular, the induced VEV for
$\phi^0_5$ allows for small mases for some spin 1/2 particles as, for
example, for the down quark $d$ and for the electron $e$ and muon $\mu$, as
we will see.

\section{\label{sec:sec4}Mass spectrum}
Masses for the particles are generated from the VEV of the
scalar fields and from the soft terms in the scalar potential.

For simplicity we assume that the VEV are real, which means 
that spontaneous CP violation through the scalar exchange is not 
considered. Now, for convenience in reading we rewrite the 
expansion of the scalar fields acquiring VEV as
\begin{eqnarray} \nonumber \label{neutras}
\phi_\tau^{'0}&=&V+ \frac{\phi_{\tau R}^{'0}+
i \phi_{\tau I}^{'0}}{\sqrt{2}}, \\ 
\phi_4^0&=&v+ \frac{\phi_{4R}^0+i \phi_{4I}^0}{\sqrt{2}}, 
\end{eqnarray}
where the subindexes $R$ and $I$ refer, respectively, to 
the real sector (CP-even scalars) and to the imaginary sector (CP-odd 
scalars or pseudoscalars) of the sleptons. 

Using Eq.~(\ref{neutras}), the minimization of the scalar potential produces the following constraints
\begin{eqnarray}\nonumber \label{minim}
\vert \lambda\vert^2-\frac{m^2_{33}}{v^2}&=&\frac{g^2_3}{6}
\left(2\frac{V^2}{v^2}-1\right) +
\frac{g^2_1}{9}\left(\frac{V^2}{v^2}-2\right) \nonumber \\
& &+\lambda^{'}_{13}\frac{\langle\phi^0_e\rangle}{2vV}, \nonumber \\
\vert \mu\vert^2+m^2_4&=&V^2\vert \lambda\vert^2+ 
\left(\frac{g^2_3}{6} +\frac{2g^2_1}{9}\right)(V^2-2v^2) \nonumber \\
& &+\lambda^{'}_{13}\frac{\langle\phi^0_e\rangle V}{2v}, \nonumber \\
m^2_{45}&=&\frac{\langle\phi^0_5\rangle}{v} \Big\{\left(V^2-2v^2\right)\left(\frac{g^2_3}{3}-\frac{4g^2_1}{9}\right)-2\vert\mu \vert^2 \nonumber \\
& & \qquad \quad -2m^2_5\Big\}, \nonumber \\
\frac{vV}{2}\lambda^{'}_{13}&=&\langle\phi^0_e\rangle\Big\{\frac{g^2_3}{6}
(V^2+v^2)-\frac{g^2_1}{9}(V^2-2v^2) \nonumber \\
& & \qquad \quad -m^2_{11}\Big\}, \nonumber \\
m^2_{13}&=&0,
\end{eqnarray}
where we have included, at the first order in $\langle\phi^0_e\rangle$ and $\langle\phi^0_5\rangle$, the radiative corrections coming from the induced VEV shown in Fig.~\ref{fig1}. Notice that by choosing $m^2_{33}<0$ and of the order of $V^2$, the parameter $\lambda$ can be as small as desired.

Our approach will be to look for consistency in the sense that the mass
spectrum must include three light spin 1/2 neutral particles (the
neutrinos) with the other spin 1/2 neutral fields having masses larger
than or equal to half of the $Z^0$ mass, to be in agreement with
experimental bounds. The consistency of the model also requires eight spin zero Goldstone bosons, four charged and four neutral ones, out of which one neutral must be related to the real sector of the sleptons (CP-even) and three neutrals to the imaginary sector (CP-odd), in order to produce masses
for the gauge bosons $W^\pm, \; K^\pm ,\; K^0, \; \bar{K}^0, \; Z^0$
and $Z^{\prime 0}$ after the breaking of the symmetry.

\subsection{\label{sec:sec41}Spectrum in the gauge boson sector}
For the SUSY version of the model the gauge bosons are the same 17 gauge 
bosons for the nonsupersymmetric version.
After breaking the symmetry with $\langle\phi_\tau\rangle +
\langle\phi_4\rangle$ and using for the covariant
derivative for triplets $iD^\mu=i\partial^\mu-(g_3/2) \lambda_\alpha
A^\mu_\alpha-g_1XB^\mu$, we get the following mass 
terms for the charged gauge bosons: 
$M^2_{W^\pm}=(g^2_3/ 2)v^2$ as in the SM, $M^2_{K^\pm}=(g^2_3/ 
2)(v^2+V^2)$, and $M^2_{K^0(\bar{K}^0)}=(g^2_3/ 2)V^2$. 
Since $W^\pm$ does not mix with $K^\pm$, and $g_2=g_3$, we have that
$v\approx 174$ GeV as in the SM.

For the neutral gauge bosons we get mass terms of the form 

\begin{eqnarray*}
M_n&=&{g_3^2\over 2}\Big\{V^2 \left(\frac{2g_1B^\mu}{3g_3}
-\frac{2A_8^\mu}{\sqrt{3}}\right)^2 \\
& &+v^2\left(A^\mu_3
+\frac{A_8^\mu}{\sqrt{3}}
-\frac{4g_1B^\mu}{3g_3}\right)^2\Big\}.
\end{eqnarray*}
 
This expression is related to a $3 \times 3$ mass matrix with a zero
eigenvalue corresponding to the photon $A^\mu_0$ given by
Eq.~(\ref{foton}). Once the photon field has been identified, we remain
with a $2 \times 2$ mass matrix for the two neutral gauge bosons $Z^\mu_0$
and $Z^{'\mu}_0$ defined in Eq.~(\ref{zetas}).
 
The physical neutral gauge bosons are defined through the mixing angle 
$\theta$ between $Z^\mu_0$ and $Z'^\mu_0$ 

\begin{eqnarray}\nonumber
Z_1^\mu&=&Z^\mu_0 \cos\theta+Z'^\mu_0 \sin\theta \; ,\\ \nonumber
Z_2^\mu&=&-Z^\mu_0 \sin\theta+Z'^\mu_0 \cos\theta, \end{eqnarray} 
where
\begin{equation} \label{tan} 
\tan(2\theta) = - \frac{v^2 \sqrt{3 - 4S^2_W}}
{2V^2 C_W^4- v^2 (2 S^2_W - 1)}, \end{equation}
with $\theta\longrightarrow 0$ in the limit $V\longrightarrow\infty$.

By using experimental results from the CERN LEP, SLAC Linear Collider and atomic parity violation data, bounds on the mass scale $V$ of the new gauge bosons and on the mixing angle $\theta$ have been calculated in Refs.~\cite{331susy,pgs,psm}. Generically, $V \geq 1$ TeV and $\theta \leq 10^{-3}$.

\subsection{\label{sec:sec42}Masses for the quark sector}
\subsubsection{\label{sec:sec421} Tree-level masses}
For the up quark sector the first two terms in the superpotential in 
Eq.~(\ref{superpotential}) produce, when we take $\langle\tilde{L}_4\rangle 
= (v,0,0)$, the following tree-level mass terms
\beqa \label{qtlu} \nonumber
{\cal L}^u_Y&=& v
(h^u_{21}u_{2L}u^{c}_{1L}+h^u_{22}u_{2L}u^{c}_{2L}+h^u_{23}u_{2L}u^{c}_{3L} 
\\ \nonumber  
&+&h^u_{31}u_{3L}u^{c}_{1L}+h^u_{32}u_{3L}u^{c}_{2L}+h^u_{33}u_{3L}
u^{c}_{3L} \\ \nonumber 
&+& h^U_2 u_{2L}U_L^c + h_3^Uu_{3L}U_L^c) + h.c.,
\eeqa
which imply bare masses only for the top $(u_3)$ and charm $(u_2)$ 
quarks. By taking $v\simeq 174$ GeV and $h^u_{22}\approx h^u_{23}\approx 
h^u_{32}\approx h^u_{33}\sim 0.5$, we can obtain appropriate values for the 
masses of the top and charm quarks, the first one of the order 
of $v$ and the second one proportional to $v$, but suppressed by the 
difference of Yukawa couplings $(h^u_{22}h^u_{33}-h^u_{23}h^u_{32})$. 
So, for the up quark sector and at the tree level, the up 
quark $u_1$ and the exotic $SU(2)_L$ singlet $U$ remain massless, even 
though their right handed components mix with the massive up type quarks. 
Later we will see how they can acquire proper masses; in special, how the 
singlet $U$ may acquire a large mass and the ordinary $u$ can acquire a 
small mass in the context of the superpotential given by 
Eq.~(\ref{superpotential}).

For the down quark sector the third and four terms in the superpotential 
produce, when we take $\langle\tilde{L}_\tau\rangle = (0,0,V)$, the following tree-level mass 
terms
\beqa \label{qtlud} \nonumber
{\cal L}^d_Y&=& V
(h^d_{213}D_{2L}d^{c}_{1L}+h^d_{223}D_{2L}d^{c}_{2L}
+h^d_{233}D_{2L}d^{c}_{3L} \\ \nonumber  
&+&h^d_{313}D_{3L}d^{c}_{1L}+h^d_{323}D_{3L}d^{c}_{2L}
+h^d_{333}D_{3L}d^{c}_{3L} \\ \nonumber 
&+& h^D_{223} D_{2L}D_{2L}^c + h^D_{233}D_{2L}D^{c}_{3L} + 
    h^D_{323} D_{3L}D_{2L}^c\\ \nonumber 
&+& h^D_{333}D_{3L}D^{c}_{3L}) + h.c.,
\eeqa
which imply bare masses of the order of $V$ for the two exotic down quarks 
and tree-level mixing of the two exotic down quarks with the right-handed 
components of the three ordinary down quarks. In what follows we are going 
to show how the ordinary down quarks can acquire proper mass values.

\subsubsection{\label{sec:sec422} One loop radiative masses}
Let us see how the quarks $U$, $b$ (bottom) and $s$ (strange) can get appropriate masses via one loop radiative corrections.

\begin{figure}
\includegraphics{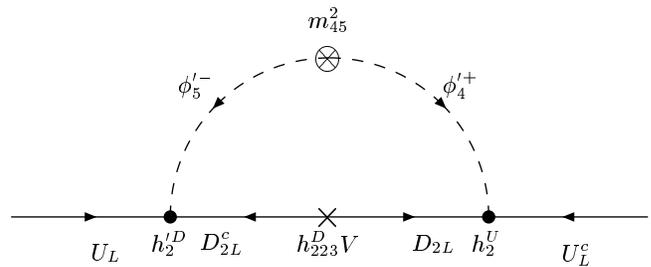}
\caption{\label{fig2}One loop diagram contributing to the radiative 
generation of the exotic up quark mass.}
\end{figure}

First, by using the Yukawa couplings in Eq.~(\ref{superpotential}) and the
parameters in $V_{\hbox{soft}}$ in Eq.~(\ref{Vsoft}) we can draw the
diagram in Fig.~\ref{fig2} which shows how the exotic quark $U$ gets a
radiative mass from the exotic down quark $D_2$. Actually there is
another diagram similar to the one in Fig.~\ref{fig2}, where $D_{3L}$ and
$D_{3L}^c$ replace $D_{2L}$ and $D_{2L}^c$, respectively, and two more
diagrams involving the squarks (leptoquarks) $\tilde{D}_{iL}$ and
$\tilde{D}^{c}_{iL}$, for $i=2,3$. Since $V\sim M_1\sim M_2 = M_{susy}$
the four diagrams are of the same order of magnitude.

Even though Fig.~\ref{fig2} is a one-loop diagram, we can expect it to
produce a mass value for the $U$ quark larger than 174 GeV (the top quark
mass) due to the fact that the mass generated is controlled by the three free, but large parameters $m_{45}^2$, $V$ and $\vert\mu\vert$, which are all at the TeV scale. In the appendix we show how a reasonable choice of the values of the parameters involved produces a large radiative mass for the exotic quark $U$.

\begin{figure}
\includegraphics{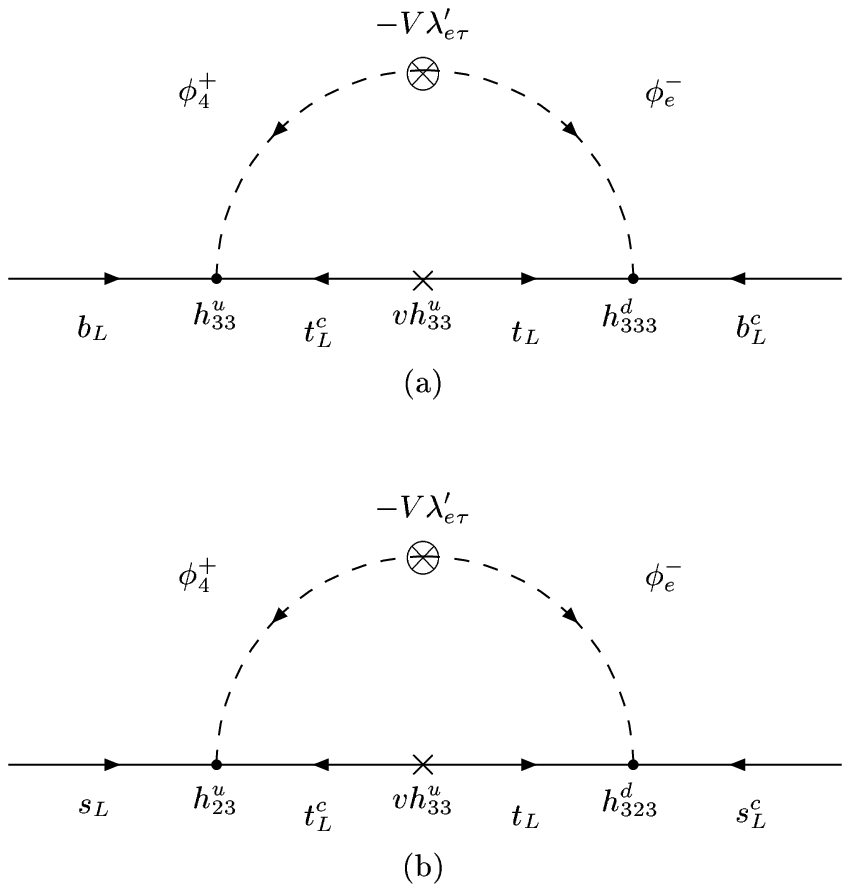}
\caption{\label{fig3}One loop diagrams contributing to the radiative 
generation of the bottom $b$ and strange $s$ quark masses.}
\end{figure}

In a similar way we show in Fig.~\ref{fig3} how the two heaviest ordinary
down quarks $b$ and $s$ get one-loop radiative masses from the top quark.  
Because the top quark mass is of the order of $v\sim 174$ GeV, these masses are at least one order of magnitude smaller than the $U$ quark mass, with the mass for the strange quark $s$ suppressed by differences of Yukawa
couplings.

Again there are four diagrams for each radiative mass, with the other
three ones involving the charm quark $c$ in the internal line, and the 
squarks $\tilde{t}$ and $\tilde{c}$.

\subsubsection{\label{sec:sec423} Higher order radiative masses}
\begin{figure}
\includegraphics{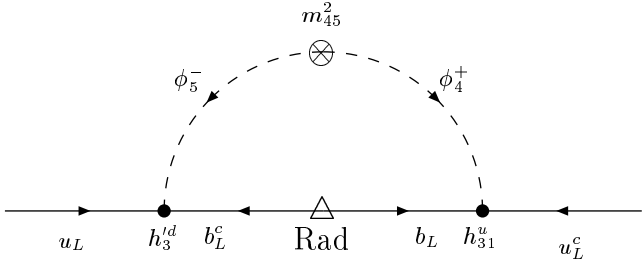}
\caption{\label{fig4}A two loop diagram contributing to the radiative 
generation of the up quark mass.}
\end{figure}

\begin{figure}
\includegraphics{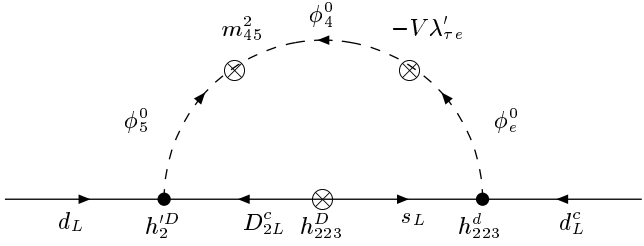}
\caption{\label{fig5}A three mixing diagram contributing to the radiative 
generation of the down quark mass.}
\end{figure}

Figs.~\ref{fig4} and \ref{fig5} show how the quarks in the first family 
acquire higher order radiative masses in the context of the 
superpotential in Eq.~(\ref{superpotential}). As a matter of fact, 
Fig.~\ref{fig4} shows how the up quark $u$ gets a second order radiative 
mass from the $b$ quark (which already has a radiatively generated mass), 
and Fig.~\ref{fig5} shows how the ordinary down quark $d$ acquires a mass 
via a triple mixing. Again, as before, the diagrams shown are not the only 
ones contributing to these masses, for example the fifth term in the 
superpotential gives a mass for the down quark of 
the form $h_1^dd_{1L}d_{1L}^c\langle\phi_5^0\rangle$ via the radiatively 
induced VEV for $\phi_5^0$ shown in Fig.~\ref{fig1}.

\subsection{\label{sec:sec43}Masses for neutralinos}
The neutralino sector is the most sensitive to the particular values of the parameters, several of which are going to be fixed in this section by the use of experimental constraints. 

For this model the neutralinos are lineal combinations of the eight neutral 
particles in $L_\alpha$ and $L_\eta$ (for $\alpha = e,\mu,\tau$ and $\eta = 
4,5$), and of the five neutral gauginos. In the basis 
$\psi = (\nu_e,\nu_\mu,\nu_\tau, N_e^0, N_\mu^0, N_\tau^0$, $N_4^0, N_5^0,
\tilde{B}^0, \tilde{A}_3^0,\tilde{A}_8^0, \tilde{K}^0, 
\tilde{\bar{K}}^0)$, the tree-level mass matrix is given by 
\begin{equation}
M=\left(\begin{array}{cc}
M_8 & M_{85} \\
M_{85}^T & M_5 \\
\end{array} \right), 
\end{equation}
where 
\begin{equation} \nonumber M_8=\left(\begin{array}{cccccccc}
0 & 0 & 0 & 0 & 0 & \lambda v/2 & \lambda V/2 & 0 \\
0 & 0 & 0 & 0 & 0 & 0 & 0 & 0 \\
0 & 0 & 0 & -\lambda v/2 & 0 & 0 & 0 & 0 \\
0 & 0 & -\lambda v/2 & 0 & 0 & 0 & 0 & 0 \\
0 & 0 & 0 & 0 & 0 & 0 & 0 & 0 \\
\lambda v/2 & 0 & 0 & 0 & 0 & 0 & 0 & 0 \\
\lambda V/2 & 0 & 0 & 0 & 0 & 0 & 0 & \mu/2 \\
0 & 0 & 0 & 0 & 0 & 0 & \mu/2 & 0 \\
\end{array}\right),
\end{equation}

\begin{equation} \nonumber 
M_{85}=\left(\begin{array}{ccccc}
0 & 0 & 0 & 0 & 0  \\
0 & 0 & 0 & 0 & 0  \\
0 & 0 & 0 & -g_3V & 0  \\
0 & 0 & 0 & 0 & 0  \\
0 & 0 & 0 & 0 & 0  \\
-\sqrt{2}g_1V/3 & 0 & 2g_3V/\sqrt{6} & 0 & 0  \\
2\sqrt{2}g_1v/3 & -g_3v/\sqrt{2} & -g_3v/\sqrt{6} & 0 & 0  \\
0 & 0 & 0 & 0 & 0  \\
\end{array} \right), 
\end{equation}
and from the soft terms in the scalar potential we read $M_5=Diag. 
(M_1,M_2,M_2,A_{2\times 2})$, where $A_{2\times 2}$ is a $2\times 2$ 
matrix with entries zero in the main diagonal and $M_2$ in the secondary 
diagonal.

This $13\times 13$ mass matrix is controlled by the parameter $\lambda$ in the sense that this parameter must be very small in order to have only three light states, with the rest of them having masses larger than half of the measured mass of the $Z^0$ neutral gauge boson. As a matter of fact, this  mass matrix has two eigenvalues equal to zero, associated with
a massless Dirac neutrino. Two more Dirac neutrinos are associated with
the values $\lambda v$ and $\mu$ and there are seven Majorana masses
different from zero, with only one of them of the see-saw type. By using $v=0.174$
TeV, $g_3 = 0.65$ and $g_1 = 0.38$, as impossed by the low energy phenomenology, we must tune the parameter $\lambda$ to lie in the range $\lambda\sim 10^{-9}$ and use for the other parameters the optimal values 
$V \sim 2$ TeV, $M_1 \sim M_2 \sim 1$ TeV, and $\vert \mu \vert \approx 10$ TeV (as we will show shortly, $\vert\mu\vert\geq 10$ TeV). With these values we obtain three light neutrinos: one Dirac neutrino with a mass of the order of the electron-Volts, one see-saw Majorana
neutrino with a mass of the order of the tenths of electron-Volts and a
zero mass Dirac neutrino (the former without including radiative
corrections which may introduce changes in this tree-level mass spectrum). All the remaining eigenvalues are above $500$ GeV.

\subsection{\label{sec:sec44}Masses for the scalar sector}
For the scalars we have three sectors, one charged and two neutrals (one 
real and the other one imaginary) which do not mix, so we can consider 
them separately.  

\subsubsection{\label{sec:sec441}The charged scalar sector}
For the charged scalars, in the basis 
$(\phi_e^-$, $\phi_\mu^-$, $\phi_\tau^-$, $\phi_e^{\prime -}$, $\phi_\mu^{\prime -}$, $\phi_\tau^{\prime -}$, $\phi_4^-$, $\phi_5^-$, $\phi_4^{\prime -}$, $\phi_5^{\prime -})$ and after using Eq.~(\ref{minim}) in the tree level approximation, we get the 
squared mass matrix $M_{cs}$ with the following nonzero entries
\begin{eqnarray} \nonumber \label{mchar}
(M_{cs})_{11}&=&\left(-\vert \lambda\vert^2-
\frac{g^2_3}{6}+\frac{g^2_1}{9}\right)V^2 \nonumber \\
& & +\left( \frac{g^2_3}{3}-\frac{2g^2_1}{9}\right)v^2+
m^2_{11}, \nonumber \\
(M_{cs})_{22}&=&m^2_{22}-\left( \frac{g^2_3}{6}-\frac{g^2_1}{9}\right) V^2 \nonumber \\
& & +\left( \frac{g^2_3}{3}-\frac{2g^2_1}{9}\right)v^2, \nonumber \\
(M_{cs})_{33}&=&\left(\vert \lambda \vert^2 + \frac{g^2_3}{2}\right) v^2, \nonumber \\
(M_{cs})_{44}&=&\vert h^e_{31}\vert^2 V^2, \nonumber \\
(M_{cs})_{66}&=&\vert h^e_{33}\vert^2 V^2, \nonumber \\
(M_{cs})_{88}&=&\vert \mu\vert^2- \left(\frac{g^2_3}{6}-
\frac{2g^2_1}{9}\right)V^2 \nonumber \\
& & - \left(\frac{g^2_3}{6}+\frac{4g^2_1}{9}\right)v^2+ m^2_5,\nonumber \\
(M_{cs})_{99}&=&\left(\vert \lambda \vert^2 + \frac{g^2_3}{2}\right) V^2,\nonumber \\
(M_{cs})_{1010}&=&\vert \mu\vert^2+ \left(\vert h^e_{31}\vert^2+
\vert h^e_{33}\vert^2+\frac{g^2_3}{3}+\frac{2g^2_1}{9}\right)V^2 \nonumber 
\\
& & -\left(\frac{g^2_3}{6}+\frac{4g^2_1}{9}\right)v^2+m^2_5, \nonumber \\
(M_{cs})_{14}&=&(M_{cs})_{41}= 2 h^e_{11}\mu v \nonumber \\
(M_{cs})_{16}&=&(M_{cs})_{61}=h^e_{13}\mu v, \nonumber \\
(M_{cs})_{18}&=&(M_{cs})_{81}=-\lambda \mu V, \nonumber \\
(M_{cs})_{34}&=&(M_{cs})_{43}=h^e_{31}\mu v,  \nonumber \\
(M_{cs})_{36}&=&(M_{cs})_{63}=h^e_{33}\mu v, \nonumber \\
(M_{cs})_{39}&=&(M_{cs})_{93}=\left(\vert \lambda \vert^2+
\frac{g^2_3}{2}\right)v V, \nonumber \\
(M_{cs})_{46}&=&(M_{cs})_{64}=h^e_{31} h^e_{33}V^2, \nonumber \\
(M_{cs})_{48}&=&(M_{cs})_{84}=\lambda h^e_{11} v V, \nonumber \\
(M_{cs})_{49}&=&(M_{cs})_{94}=h^e_{31} \mu V, \nonumber \\
(M_{cs})_{410}&=&(M_{cs})_{104}=\frac{1}{2}h^{e\prime}_{13} V, \nonumber \\
(M_{cs})_{68}&=&(M_{cs})_{86}=\lambda h^e_{13} v V, \nonumber \\
(M_{cs})_{69}&=&(M_{cs})_{96}=h^e_{33} \mu V \nonumber \\
(M_{cs})_{610}&=&(M_{cs})_{106}=\frac{1}{2}h^{e\prime}_{33} V.
\end{eqnarray}
This mass matrix has two zero eigenvalues which correspond to the four Goldstone bosons needed to give masses to the gauge bosons $W^\pm$ and $K^\pm$. By using $h^e_{11} =h^e_{13} =h^e_{31} =h^e_{33} = 1$, $h^{e\prime}_{13} = h^{e\prime}_{33} = 1$ GeV, $m^2_{11}=m^2_{22}=m^2_5 = -m^2_{33}=1$ TeV$^2$, and with the numerical values for the other parameters as stated before, we obtain that all the nonzero eigenvalues are above $850$ GeV and so, contrary to other models, there are not charged scalars at the electroweak scale in the model analyzed here, which is something
expected due to the fact that the members of the isospin doublet in
$\phi_4$, which are absorbed by $W^\pm_\mu$, are the only charged scalars
available at the electroweak scale. This result is in agreement with the
so-called ``extended survival hypothesis" \cite{esh} which consists in
assuming that the components of the Higgs representations required for the
breaking of a particular symmetry are the only ones which are not
superheavy (``scalar Higgs fields acquire the maximum mass compatible with
the pattern of symmetry breaking" \cite{esh}).

\subsubsection{\label{sec:sec442}The neutral scalar sector}
For the neutral CP-even scalars, in the basis 
$(\phi^0_{eR}, \phi^0_{\mu R}, \phi^0_{\tau R}, \phi^{'0}_{eR},
\phi^{'0}_{\mu R}, 
\phi^{'0}_{\tau R}, \phi^0_{4 R}, \phi^0_{5 R})$ and after using 
Eq.~(\ref{minim}) in the tree level approximation, we get the squared mass matrix $M_{e}$ with the following nonzero entries
\begin{eqnarray} \nonumber \label{even}
(M_{e})_{11}&=&\frac{1}{2}\Big\{ m^2_{11}-\left(\frac{g^2_3}{6}-
\frac{g^2_1}{9}+\vert \lambda \vert^2 \right)V^2 \nonumber \\
& & -\left(\frac{g^2_3}{6}+\frac{2g^2_1}{9}+\vert \lambda 
\vert^2\right)v^2 
\Big\}, \nonumber \\
(M_{e})_{22}&=&\frac{1}{2}\Big\{ m^2_{22}-\left(\frac{g^2_3}{6}-
\frac{g^2_1}{9} \right)V^2 \nonumber \\
& & -\left(\frac{g^2_3}{6}+\frac{2g^2_1}{9}\right)v^2 \Big\}, \nonumber \\
(M_{e})_{44}&=&\frac{1}{2} \left(  m^2_{11}- m^2_{33}\right), \nonumber \\
(M_{e})_{55}&=& \frac{1}{2}\Big\{ m^2_{22}+\left(\frac{g^2_3}{3}+
\frac{g^2_1}{9} \right)V^2 \nonumber \\
& & -\left(\frac{g^2_3}{6}+\frac{2g^2_1}{9}\right)v^2 \Big\}, \nonumber \\
(M_{e})_{66}&=&\left(\frac{g^2_3}{3}+\frac{g^2_1}{9}\right)V^2, \nonumber 
\\
(M_{e})_{77}&=&\left(\frac{g^2_3}{3}+\frac{4g^2_1}{9}\right)v^2, \nonumber 
\\
(M_{e})_{88}&=&\frac{1}{2}\Big\{\vert \mu \vert^2+ m^2_{5}-
\left(\frac{g^2_3}{6}-\frac{2g^2_1}{9} \right)V^2 \nonumber \\
& & +\left(\frac{g^2_3}{3}-\frac{4g^2_1}{9}\right)v^2 \Big\}, \nonumber \\
(M_{e})_{18}&=&(M_{ch})_{81}=\frac{\lambda}{2}\mu V, \nonumber \\
(M_{e})_{67}&=&(M_{ch})_{76}=-\left( \vert \lambda \vert^2 + 
\frac{g^2_3}{6}+\frac{2g^2_1}{9}\right)v V. \nonumber \\ 
\end{eqnarray}
The matrix $M_{e}$ has one eigenvalue equal to zero, corresponding to
one Goldstone boson. The nonzero eigenvalues, which can be calculated analitically, are 
\begin{eqnarray}\nonumber
m^2_{e1}&=& (M_{e})_{22}, \nonumber \\
m^2_{e2}&=& (M_{e})_{44}, \nonumber \\
m^2_{e3}&=& (M_{e})_{55}, \nonumber \\
m^2_{e4}&=& \frac{1}{2}\Big\{(M_{e})_{11}+(M_{e})_{88} \nonumber \\
& & +\sqrt{[(M_{e})_{11}-(M_{e})_{88}]^2+4[(M_{e})_{18}]^2}\Big\}, \nonumber \\
m^2_{e5}&=& \frac{1}{2}\Big\{(M_{e})_{11}+(M_{e})_{88} \nonumber \\
& & -\sqrt{[(M_{e})_{11}-(M_{e})_{88}]^2+4[(M_{e})_{18}]^2}\Big\}, \nonumber \\
m^2_{e6}&=& \frac{1}{2}\Big\{(M_{e})_{66}+(M_{e})_{77} \nonumber \\
& & +\sqrt{[(M_{e})_{66}-(M_{e})_{77}]^2+4[(M_{e})_{67}]^2}\Big\}, \nonumber \\
m^2_{e7}&=& \frac{1}{2}\Big\{(M_{e})_{66}+(M_{e})_{77} \nonumber \\
& & -\sqrt{[(M_{e})_{66}-(M_{e})_{77}]^2+4[(M_{e})_{67}]^2}\Big\},
\end{eqnarray}
where $m_{e7}$ is associated with the lightest CP-even Higgs scalar $h$. By introducing the numerical values for the parameters as calculated in the previous sections, we obtain $m_{e7}=m_h\approx 85$ GeV which is a bit larger than the lowest bound on the lightest CP-even Higgs scalar in the MSSM \cite{mssm}. All the remaining eigenvalues are above $750$ GeV.. Notice that the scalar $h$ is a mixture of $\phi_{4R}^0$ and $\phi_{\tau R}^{\prime 0}$, as it should be according to the ``extended survival hypothesis" \cite{esh}, and because of this it partially decouples from the $Z^0$ of the SM at high energies, since it is a mixture of a singlet and a doublet under $SU(2)_L$, with the singlet having an $U(1)_Y$ hypercharge equal to zero.

For the neutral CP-odd scalars, in the basis 
$(\phi^0_{eI}, \phi^0_{\mu I}, \phi^0_{\tau I}, \phi^{'0}_{eI},
\phi^{'0}_{\mu I}, 
\phi^{'0}_{\tau I}, \phi^0_{4 I}, \phi^0_{5 I})$ and after using 
Eq.~(\ref{minim}) in the tree level approximation, we get the squared mass matrix $M_{o}$ with the following nonzero entries

\begin{eqnarray} \nonumber \label{odd}
(M_{o})_{11}&=&(M_{e})_{11}, \nonumber \\
(M_{o})_{22}&=&(M_{e})_{22}, \nonumber \\
(M_{o})_{44}&=&(M_{e})_{44}, \nonumber \\
(M_{o})_{55}&=&(M_{e})_{55}, \nonumber \\
(M_{o})_{88}&=&(M_{e})_{88}, \nonumber \\
(M_{o})_{18}&=&(M_{o})_{81}=(M_{e})_{18}. 
\end{eqnarray}
This mass matrix has three zero eigenvalues, which correspond to three additional Goldstone bosons. The five nonzero eigenvalues, at the tree level, are 
\begin{eqnarray} \nonumber
m^2_{o1}&=&m^2_{e1}, \qquad m^2_{o2}=m^2_{e2}, \nonumber \\
m^2_{o3}&=&m^2_{e3}, \qquad m^2_{o4}=m^2_{e4}, \nonumber \\
m^2_{o5}&=&m^2_{e5},
\end{eqnarray}
equal to five of the eigenvalues in the real sector as a consequence of our assumption that there is not CP violation in the neutral scalar sector. Notice, by the way, that this model does not have a light pseudoescalar particle.

The four Goldstone bosons associated to the neutral scalar sector will
provide masses for the gauge bosons $K^0_\mu$, $\bar{K}^0_\mu$, $Z^0_\mu$,
and $Z^{'0}_\mu$.
  
\subsection{\label{sec:sec45}Masses for charginos}
\begin{figure}
\includegraphics{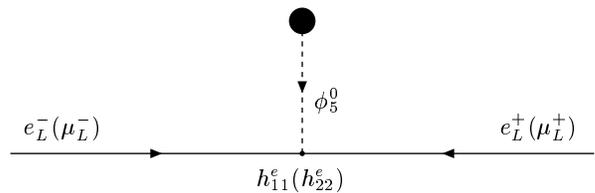}
\caption{\label{fig6}Radiatively induced VEV contributing to the $e$ and 
$\mu$ masses.}
\end{figure}

The charginos in the model are lineal combinations of the ordinary charged 
leptons, the two exotic electrons and the two charged gauginos. In the 
gauge eigenstate basis 
$\psi^\pm = (e^+, \mu^+, \tau^+$, $E_4^+, E_5^+,\tilde{W}^+, \tilde{K}^+, 
e^-, \mu^-, \tau^-, E_4^-, E_5^-,\tilde{W}^-, \tilde{K}^-)$ the tree-level 
chargino mass terms in the Lagrangian are of the form 
$\psi^\pm M_{ch}\psi^\pm$ where 
\begin{equation}
M_{ch}=\left(\begin{array}{cc}
0 & M_7 \\
M_7^T & 0 \\
\end{array} \right), 
\end{equation}
and  
\begin{equation}
M_7=\left(\begin{array}{ccccccc}
0 & 0 & 0 & 0 & h^e_{31}V & 0 & 0 \\
0 & 0 & 0 & 0 & 0         & 0 & 0 \\
0 & 0 & 0 & 0 & h^e_{33}V & 0 & 0 \\
-\lambda V& 0 & 0 & \mu & 0 & -g_3v & 0 \\
0 & 0 & 0 & 0 & \mu & 0 & -g_3v \\
0 & 0 & 0 & 0 & 0 & M_2 & 0 \\
0 & 0 & -g_3V & 0 & h^e_{31}V & 0 & M_2 \\
\end{array} \right).
\end{equation}
This mass matrix has two eigenvalues of the order of the $\mu$ scale, 
two other eigenvalues of the order of the SUSY scale ($M_2\sim 1$ 
TeV), two eigenvalues equal to zero, and one see-saw eigenvalue of the 
order of $2M_2^2v^2/\mu^2$, which for $\mu\gg M_2\sim V$ can account for 
the tau lepton mass. Notice that the tau mass is related to the mass parameter $M_2$ of the corresponding gaugino, but it is suppressed by the parameter $\mu$. So, at the tree-level only the electron $e$ and the muon $\mu$  remain massless, but they pick up radiative masses . In fact, the seventh term in the superpotential produces immediately the diagram in Fig.~\ref{fig6} which shows how both, $e^-$ and $\mu^-$, get finite masses via the radiatively induced VEV for the scalar field $\phi_5^0$ shown in 
Fig.~\ref{fig1}. (This mechanism has been used in the literature in 
Ref.~\cite{liu} in order to generate masses for charged fermions. See also 
Ref.~\cite{babu}.)

\section{\label{sec:sec5}General remarks and conclusions} 
We have built the complete supersymmetric version of a 3-3-1 model for
three families, the simplest one we have been able to imagine. Contrary to
the MSSM which has two Higgs doublets at the electroweak energy scale, in
this model there is only one $SU(2)_L$ Higgs doublet acquiring a non-zero
VEV (the one associated to $\phi_4$). So, the MSSM is not an effective theory of the model we have constructed.

For the model presented in this paper the slepton multiplets play the role 
of the Higgs scalars and no Higgsinos are required, which implies a 
reduction of the number of parameters and degrees of freedom, compared 
to other 3-3-1 supersymmetric models in the literature \cite{ma}. 

For our analysis we have taken the simplest VEV possible, able to break the
symmetry and give, at the same time, masses to the fermion fields in the 
model. The choice of this simple VEV structure was dictated not only
by simplicity, but also by paying due attention to the general mass
spectrum of the particles. The most general VEV structure for this model is of the form:
$\langle\phi_e\rangle = (0, v_e, V_e), 
\langle\phi_\mu\rangle = (0, v_\mu, V_\mu), 
\langle\phi_\tau\rangle = (0, v_\tau, V_\tau), 
\langle\phi_4\rangle = (v_4, 0, 0)$ and  
$\langle\phi_5\rangle = (v_5, 0, 0)$, which, even when properly 
rotated $(v_e=V_\mu =0)$, gives a very messy scalar sector and can dramatically change  
the mass spectrum presented here. Obviously, there are in this general VEV 
structure a lot of free parameters to play with, some of them able to 
solve possible inconsistencies in the mass spectrum calculated. 

There are in the model three mass scales: The electroweak scale
$v\approx 174$ GeV (a value dictated by the weak $W^\pm$ gauge boson mass), which is the same mass scale associated with the SM;  
the SUSY-(3-3-1) mass scale $M_1\sim M_2 \sim V\sim 1-2$ TeV, which is the
same scale associated with the MSSM; and the $\mu$ scale which can be as
large as the Planck scale, but whose value is fixed (by the tau lepton mass)
to lie in the range 10 TeV $\leq \mu \leq $ 100 TeV. As can be realized,
for this model we have the same $\mu$ problem that is present in the MSSM,
and it should find an explanation outside the context of the analysis
presented here.

We have aligned the vacuum in the way enunciated in the main text,
inspired in the non-SUSY model presented in Sec.~\ref{sec:sec2}; this
alignment produces a consistent mass spectrum for quarks and charged
leptons in the following way: First, the exotic down quarks and leptons
get masses of the order of the SUSY scale; then the top and charm quarks
get tree-level masses at the electroweak energy scale, with the mass for the charm quark suppressed by differences of Yukawa couplings. The exotic up quark $U$ gets a one-loop radiative mass which, for specific values of the Yukawa couplings, can be made larger than 174 GeV, the top quark mass (see the appendix). The bottom quark and the strange quark get one loop radiative masses, with the mass of the strange quark suppressed by
differences of Yukawa couplings; then the up quark and the down quark get higher order radiative masses. For the known charged leptons only the tau gets a tree-level mass at the electroweak scale, but suppressed by a see-saw mechanism, with the electron and muon acquiring loop masses via
radiatively induced VEV. The analysis for the neutrino sector has not been
completed yet, but the preliminary analysis presented in the main text
does not show inconsistencies.

The vector-like structure $L_{4L}\oplus L_{5L}$, which seems irrelevant
for the non-SUSY model presented in Sec.~\ref{sec:sec2} because it does
not contribute to the anomaly constraint equations, is mandatory in this
SUSY version of the model because without its presence it is not possible to provide with masses for the lepton sector.

The scalar sector of the model presented here is so rich that even with
the simple VEV used, it is able to reproduce a consistent mass spectrum
for the spin 1/2 particles by using different radiative mechanisms, some
of them new. For example, the way how the mass for the down quark
$d$ is generated via a three-mixing loop diagram plus a radiatively induced
VEV has not been used in the literature yet, as far as we know.

Even though the algebra involved in all the equations related with the
scalar sector (Secs.~\ref{sec:sec32} and \ref{sec:sec44}) is quite
tedious, the final results are simple, with mass matrices that admit
analytic solutions and neat physical interpretations. This is just a
byproduct of the $Z_2$ symmetry introduced in Sec.~\ref{sec:sec31},
whose reason of being is the suppression of proton decay, with its final form dictated by discrete anomaly cancellation constraint relationships, and by the mass spectrum for the lepton fields.

In the main text we have also calculated the mass value, at the tree level,
of the lightest CP-even Higgs scalar $h$ which is larger than the
lowest bound on the lightest CP-even Higgs scalar in the MSSM, in spite of being a mixture between a member of a pure $SU(2)_L$ doublet $\phi^0_{4R}$ and the singlet $\phi^{\prime 0}_{\tau R}$.

\begin{figure}
\includegraphics{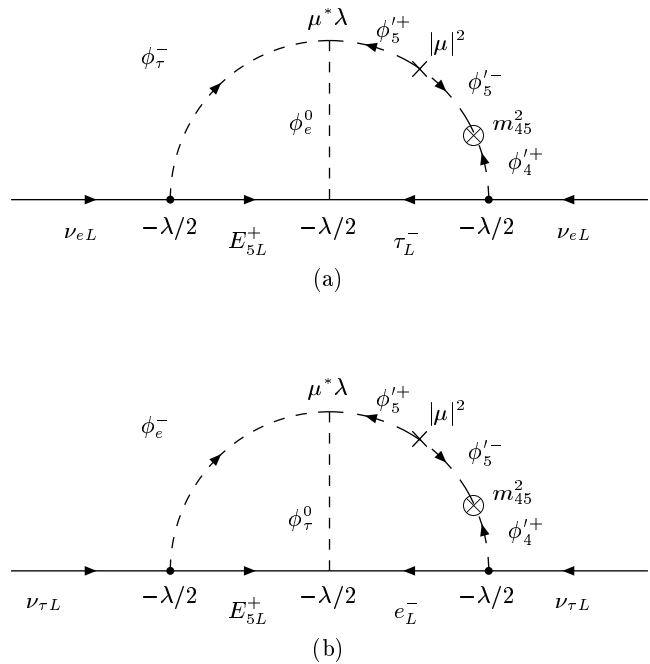}
\caption{\label{fig7}Zee mechanism for neutrino masses.}
\end{figure}

To conclude, let us say that we feel a little unpleasant with the small
value $\lambda\sim 10^{-9}$, which seems unnatural and may require some
sort of fine tuning. We can avoid this inconvenience by letting $\phi_5$,
instead of $\phi_4$, to acquire the zero order VEV $(v,0,0)$. Then no
tree-level mass terms for the neutrinos show up, but two of the neutrinos
do acquire a two-loop radiative mass via a kind of Zee mechanism
\cite{zee} as depicted in Fig.~\ref{fig7}, which are two among other
graphs, and show that the Zee mechanism is automatically present in the
model discussed here. The price we have to pay if we want to use this
mechanism in order to generate neutrino masses is the explanation of the
mass spectrum for the entire up quark sector which must be generated via
radiative corrections. In any case, the mass scale for the neutrinos is
controlled by the parameter $\lambda$ which is not present in the quark
sector of the model.

\section*{ACKNOWLEDGMENTS}
We acknowledge partial financial support from Universidad de Antioquia. L. A. S. acknowledges partial financial support from DIME at the Universidad Nacional-Sede Medell\'\i n.  W. A. P. thanks the Theoretical Physics Laboratory at the Universidad de La Plata in Argentina where part
of this work was done.

\appendix*
\section{}
In this appendix we calculate the diagram in Fig.~\ref{fig2} and analyze its numerical value. The algebra shows that this diagram is finite and proportional to

\begin{eqnarray}\label{diag1} \nonumber
\Delta &=&N[M^2m_{c4}^2\log(M^2/m_{c4}^2) - M^2m_{c5}^2\log(M^2/m_{c5}^2)\\
&+&m^2_{c4}m_{c5}^2\log(m^2_{c4}/m_{c5}^2)],
\end{eqnarray}
where $N =h_2^{\prime D}h_2^U m_{45}^2M/[16\pi^2(m_{c5}^2-m_{c4}^2)
(M^2-m_{c4}^2)(M^2-m_{c5}^2)],\; M=h_{223}^DV$ is the mass of the heavy exotic down quark $D$, 
and $m_{c4}$ and $m_{c5}$ are the masses of $\phi_4^{\prime +}$ and $\phi_5^{\prime -}$, 
respectively. To estimate a value for $\Delta$ we use the following values obtained in Sects. 4.3 and 4.4 of this paper: $m_{c4}\approx (M_{cs})_{99}\approx \sqrt{g^2_3/2}V, \; m_{c5}\approx (M_{cs})_{1010}\approx |\mu|\approx 10$ TeV, and 
$m_{45}^2\approx -2\langle\phi_5^0\rangle |\mu|^2/v$. Notice that the value of $\Delta$ is a function of the dimensionless parameter $\langle\phi_5^0\rangle /v$ and of the Yukawa couplings $h_{223}^D$, $h_2^{\prime D}$ and $h_2^U$. We are going to put the three Yukawa constants equal to a common value $h$. The point now is to assign values to $h$ and also to the radiative correction $\langle\phi_5^0\rangle /v$ which, being a radiative correction to scalar masses in a supersymmetric model, can be large. Table \ref{tab1} shows the numerical evaluation of $\Delta$ (the mass of the exotic quark $U$) as a function of these parameters. 
\begin{table}[h]
\caption{\label{tab1}Radiative mass $\Delta$ for the exotic quark $U$}
\begin{ruledtabular}
\begin{tabular}{ccc}
$\langle\phi_5^0\rangle /v$ & $h$ & $\Delta$ (GeV)\\
\hline
$0.1$ & $4.1$ & $203.5$ \\
$0.2$ & $3.0$ & $207.8$ \\
$0.4$ & $2.2$ & $206.0$ \\
$0.6$ & $1.9$ & $219.2$ \\
\end{tabular}
\end{ruledtabular}
\end{table}

So, by a reasonable choice of the values of the parameters involved in the calculation of the diagram in Fig.~\ref{fig2}, we can obtain a radiative mass for the exotic up quark $U$ larger than the top quark mass.

{\it Note added in proof:} After the completion of the first draft of this manuscript, we became aware of the existence of a related paper by J.C. Montero, V. Pleitez and M.C. Rodriguez entitled ``Supersymmetric 3-3-1 model with right-handed neutrinos" \cite{mpr}. Even though the gauge and quark sectors are the same in the two papers, they differ in the lepton and scalar sectors due to the fact that in our model we introduce the vector-like structure $L_{4L}\oplus L_{5L}$. As a consequence, and contrary to the Montero, Pleitez and Rodriguez analysis, we avoid the introduction of Higgssinos in our study. Because of this the results are different in the two papers, conspicuously enough in the three scalar sectors.


\begin{thebibliography}{}

\bibitem[1]{sm}
For an excellent compendium of the SM see: J.F. Donoghue, E. Golowich, and 
B. Holstein, \textit{``Dynamics of the Standard Model"}, (Cambridge  
University Press, Cambridge, England, 1992).

\bibitem[2]{lep2}
LEPEWWG, hep-ex/0112021, home page: http://lepewwg.web.cern.ch/LEPEWWG/.

\bibitem[3]{mssm}
H.E. Haber and G.L. Kane, Phys. Rep. \textbf{117}, 75 (1985).

\bibitem[4]{neutrinos}
For recent reviews, see: J.W.F. Valle, {\it Neutrino masses twenty-five 
years later}, hep-ph/0307192; V. Barger, D. Marfalia and K. Whishnaut, Int. J. Mod. Phys. E \textbf{12}, 569 (2003).

\bibitem[5]{fp}
F. Pisano and V. Pleitez, Phys. Rev. D \textbf{46}, 410 (1992);  P.H.
Frampton, Phys. Rev. Lett. \textbf{69}, 2887 (1992); J.C. Montero, F.
Pisano and V. Pleitez, Phys. Rev. D \textbf{47}, 2918 (1993); V. Pleitez
and M.D. Tonasse, Phys. Rev. D \textbf{48}, 2353 (1993); \textit{ibid}
5274 (1993);  D. Ng, Phys. Rev. D \textbf{49}, 4805 (1994); L. Epele, H.
Fanchiotti, C. Garc\'\i a Canal and D. G\'omez Dumm, Phys. Lett. B
\textbf{343} 291 (1995);  M. \"Ozer, Phys. Rev. D \textbf{54}, 4561
(1996).

\bibitem[6]{long}
M. Singer, J.W.F. Valle and J. Schechter, Phys. Rev. D \textbf{22}, 738
(1980);  R. Foot, H.N. Long and T.A. Tran, Phys. Rev. D \textbf{50}, R34
(1994); H.N. Long, Phys. Rev. D \textbf{53}, 437 (1996); \textit{ibid} 
\textbf{54}, 4691 (1996); V. Pleitez, Phys. Rev. D \textbf{53}, 514 (1996).

\bibitem[7]{ponce}
W.A. Ponce, J.B. Fl\'orez and L.A. S\'anchez, Int. J. Mod. Phys. A \textbf{
17}, 643 (2002); W.A. Ponce, Y. Giraldo and L.A. S\'anchez, 
\textit{``Systematic study of 3-3-1 models"}, in proceedings of the VIII 
Mexican Workshop of particles and fields, Zacatecas, Mexico, 2001. Edited 
by J.L. D\i az-Cruz \textit{et al}. (AIP Conf. Proceed. Vol. \textbf{623}, 
N.Y., 2002), pp. 341-346.

\bibitem[8]{ozer}
M. \"Ozer, Phys. Rev. D \textbf{54}, 1143, (1996); 
T. Kitabayashi, \textit{ibid} \textbf{64}, 057301, (2001).

\bibitem[9]{331susy}
J.M. Mira, W.A. Ponce, D.A. Restrepo and L.A. S\'anchez, Phys. Rev. D 
\textbf{67}, 075002 (2003); R. Mart\'\i nez, N. Poveda and J.-Alexis 
Rodr\'\i guez, Phys. Rev. D \textbf{69}, 075013 (2004).

\bibitem[10]{ma}
T.V. Duong and E. Ma, Phys. Lett. B \textbf{316}, 307 (1993); H.N. Long and
P.B. Pal, Mod. Phys. Lett. A \textbf{13}, 2355 (1998); J.C. Montero, V.
Pleitez and M.C. Rodriguez, Phys. Rev. D \textbf{65}, 035006; 095008
(2002); M. Capdequi-Peyren\`ere and M.C. Rodriguez, Phys. Rev. D
\textbf{65}, 035001 (2002).

\bibitem[11]{del}
R. Delbourgo, Phys. Lett. B \textbf{40}, 381 (1972); L. Alvarez-Gaume, Nucl. Phys. B \textbf{234}, 262 (1984).

\bibitem[12]{pgs}
W.A. Ponce, Y. Giraldo and L.A. S\'anchez, Phys. Rev. D \textbf{67}, 075001 
(2003).

\bibitem[13]{sher}
J.-Alexis Rodr\'\i guez and M. Sher, Phys. Rev. D \textbf{70}, 117702 
(2004), and references therein.

\bibitem[14]{KW}
L.M. Krauss and F. Wilczek, Phys. Rev. Lett. \textbf{62}, 1221 (1989);
L.E. Iba\~nez and G.G. Ross, Phys. Lett. B \textbf{260}, 291 (1991).

\bibitem[15]{psm}
L.A. S\'anchez, W.A. Ponce and R. Mart\'\i nez, Phys. Rev. D \textbf{64}, 075013 (2001); R. Mart\'\i nez, W.A. Ponce and L.A. S\'anchez, Phys. Rev. D \textbf{65}, 055013 (2002).

\bibitem[16]{esh}
F. del Aguila and L. Iba\~nez, Nucl. Phys. B \textbf{177}, 60 (1981).

\bibitem[17]{liu}
D. Chang and J. Liu, Phys. Rev. D \textbf{38}, 327 (1988); M. Olechowski, 
Z. Phys. C \textbf{37}, 401 (1988).

\bibitem[18]{babu}
K.S. Babu and E. Ma, Mod. Phys. Lett. A \textbf{20}, 1975 (1989).

\bibitem[19]{zee}
A. Zee, Phys. Lett. B \textbf{93}, 389 (1980); {\it ibid} B \textbf{161}, 141 (1985). For a recent discussion see D. Chang and A. Zee, Phys. Rev. D \textbf{61}, 071303 (2000).

\bibitem[20]{mpr}
J.C. Montero, V. Pleitez and M.C. Rodriguez, Phys. Rev. D \textbf{70}, 075004 (2004).

\end{thebibliography}
\end{document}